\documentclass[aps,prl,twocolumn,showpacs,floatfix,superscriptaddress,tightenlines,amsmath,amssymb,yhmath]{revtex4-2}
\usepackage{hyperref}
\usepackage[dvipsnames,x11names]{xcolor}
\pagecolor{white}
\usepackage[final]{graphicx}
\usepackage{bm}
\usepackage{float}
\usepackage{enumerate, enumitem}
\usepackage[normalem]{ulem}
\usepackage[capitalize]{cleveref}

\graphicspath{{figs_nlu/}}

\newcommand{\dd}{\mathop{}\!\mathrm{d}}

\newcommand{\vel}{\bm{u}}
\newcommand{\pol}{\bm{p}}
\newcommand{\KE}{{\mathcal K}}
\newcommand{\Rey}{{\mathrm{Re}}}

\newcommand{\bdel}{\bm{\nabla}}

\newcommand{\lp}{\lambda_+}
\newcommand{\lm}{\lambda_{-}}

\newcommand{\h}{\bm{h}}

\newcommand{\Sa}{\bm{\Sigma}^{a}}
\newcommand{\sa}{\sigma_{a}}
\newcommand{\Sij}{\bm{S}}
\newcommand{\Om}{\bm{\Omega}}

\newcommand{\lE}{\ell_{E}}
\newcommand{\avgp}{\left|\left<\bm{p}\right>\right|}

\newcommand{\Ep}{\mathcal{E}_{\pol}(q)}
\newcommand{\Ec}{\mathcal{E}_{c}(q)}
\newcommand{\Eu}{\mathcal{E}_{\vel}(q)}

\begin{document}

\title{Inertia drives concentration-wave turbulence in swimmer suspensions}
\author{Purnima Jain}
\affiliation{Tata Institute of Fundamental Research, 36/P, Gopanpally Village, Serilingampally Mandal, Ranga Reddy District, Hyderabad 500046, Telangana, India}
\author{Navdeep Rana}
\affiliation{Max Planck Institute for Dynamics and Self-Organization (MPIDS), D-37077 G\"ottingen, Germany}
\author{Sriram Ramaswamy}
\affiliation{Centre for Condensed Matter Theory, Department of Physics,Indian Institute of Science, Bangalore 560012, India}
\affiliation{International Centre for Theoretical Sciences, Tata Institute of Fundamental Research, Bangalore 560089 India}
\author{Prasad Perlekar}
\email{perlekar@tifrh.res.in}
\affiliation{Tata Institute of Fundamental Research, 36/P, Gopanpally Village, Serilingampally Mandal, Ranga Reddy District, Hyderabad 500046, Telangana, India}

\begin{abstract}
    We discover an instability mechanism in suspensions of self-propelled particles that does not involve active stress. Instead, it is driven by a subtle interplay of inertia, swimmer motility, and concentration fluctuations, through a crucial time lag between the velocity and the concentration field. The resulting time-persistent state seen in our high-resolution numerical simulations consists of self-sustained waves of concentration and orientation, transiting from regular oscillations to wave turbulence. 
    We analyze the statistical features of this active turbulence, including an intriguing connection to the Batchelor spectrum of passive scalars.
\end{abstract}

\maketitle

Hydrodynamic theories of active matter \cite{ramaswamy2010,marchetti2013,ramaswamy2019,alert2022} are remarkably successful in predicting and accounting for the complex spatiotemporal flows arising in suspensions of motile organisms. Swimming is intrinsically force-free \cite{batchelor1970}: the total force on swimmer plus fluid is zero, so the resulting force density on the suspension must have zero monopole moment.
A general swimmer is thus minimally characterized by two \textit{a priori} independent parameters -- a swimming speed $v_0$ and a force dipole of strength $W$. An aligned collection of swimmers with concentration $c$ has a force-dipole density of magnitude $Wc \equiv \sigma_a$ -- the active stress \cite{simha2002} -- which, for a system with viscosity $\mu$ and negligible inertia, as in microbial systems, drives a seemingly inexorable instability of aligned states, with exponential growth rate $\sigma_a/\mu$ \cite{simha2002}, culminating in active turbulence \cite{thampi2013velocity,alert2022}. A solid substrate~\cite{maitra2020,sarkar2021} or, more subtly, a fluid interface~\cite{maitra2023}, mitigate this instability. An orientation field is not essential for active turbulence; see \cite{padhan2024novel}.

Studies focusing on mesoscale swimmers have made a strong case for investigating active suspensions in realms where inertial effects are significant \cite{wang2012, klotsa2019,chatterjee2021,koch2021,rorai2022, derr_2022, rana2024} although distinct from the inviscid regime explored, e.g., in \cite{filella2018}.
We showed recently \cite{chatterjee2021, rana2024} that in such a suspension with mass density $\rho$, when concentration fluctuations are ignored,
the stability of a globally aligned state of extensile ($W>0$) swimmers depends on the dimensionless combination $R=\rho v_0^2/2\sa$ -- the square of the ratio of $v_0$ to the speed $\sqrt{\sa/\rho}$ at which the bend-instability would invade the system \cite{simha2002,chatterjee2021} if $v_0$ were zero, i.e., with active stress but no macroscopic motility. The bend instability in the Stokesian limit $R=0$ survives up to a threshold $R=R_2$,
beyond which the ordered suspension is linearly stable \cite{chatterjee2021,rana2024}. 
For moderate Reynolds number ($\Rey$) at the scale of the swimmer size $d$, the viscous estimate $W\sim \mu v_0 d^2$ \cite{lauga2009,chatterjee2021,derr_2022} should hold.
In that case $R\sim \phi^{-1}\Rey$ where the volume fraction $\phi = c d^3$ should be taken to be of $\mathcal{O}(1)$ as we are considering spontaneously aligned states. Thus, a modest $\Rey$ suffices for a flock in a fluid to outswim and escape its Stokesian instability \cite{chatterjee2021}.

\begin{figure}[!t]
    \centering
   \includegraphics[width=0.4\textwidth]{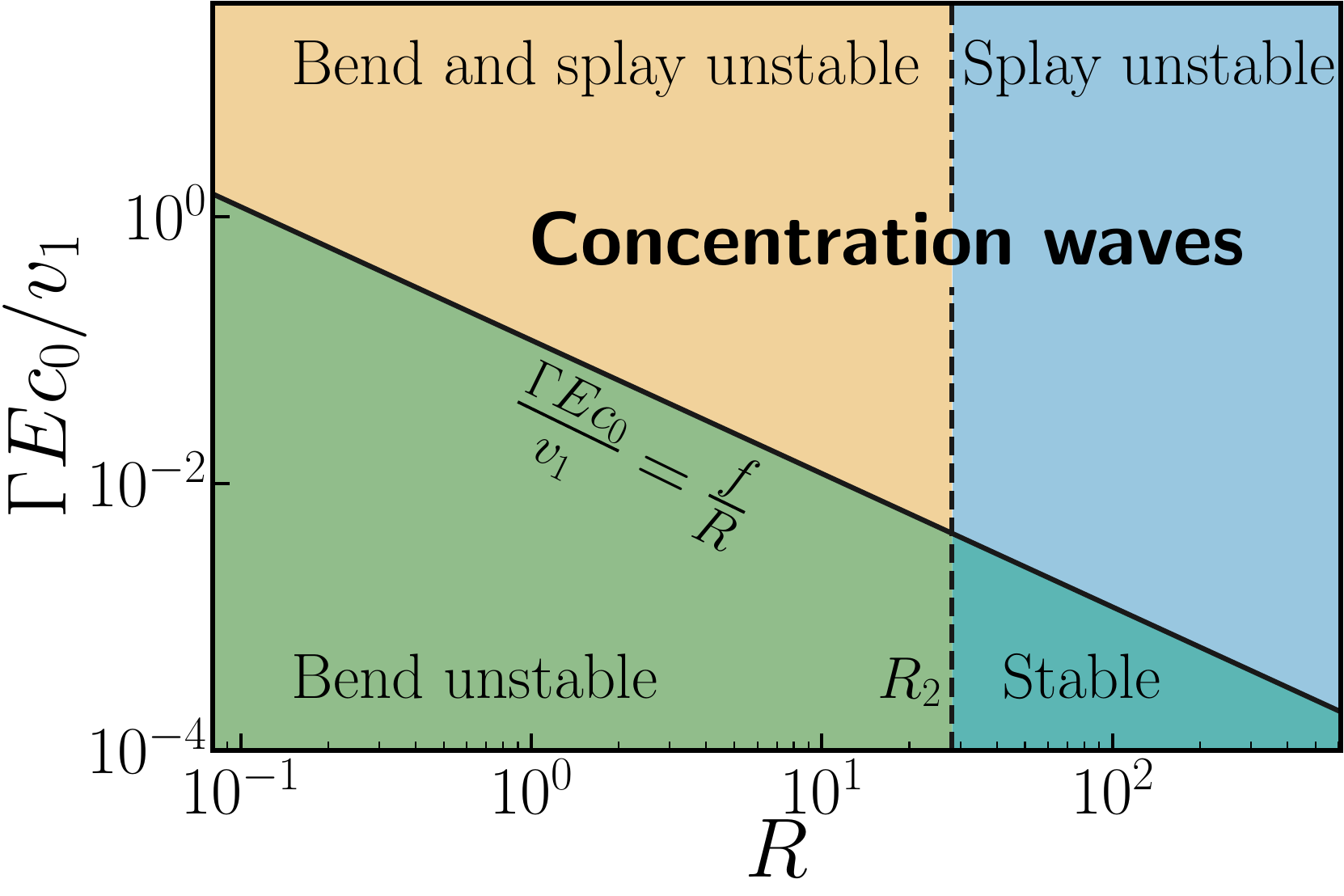}
    \caption{\label{fig:phase-diag} Representative linear stability diagram for $\rho=1$, $\mu=0.1$, $v_0=v_1=0.1$, $\lambda=0.1$, $\Gamma=1$, $K=10^{-3}$, $D=10^{-4}$, $p_0 = \sqrt{0.1}$, and $c_0=1$ (parameters in \eqref{eq:conc}-\eqref{eq:vel}). Vertical dashed line marks the stability boundary $R=R_2$ for bend perturbations \cite{chatterjee2021}. Solid black line is the novel splay-concentration stability boundary $\Gamma E c_0/v_1 = f/R$ \eqref{eq:threshold_E} \cite{Note1}.} 
\end{figure}

In this Letter, we uncover a manifestation of the roles of self-propulsion \textit{and} inertia with a character qualitatively distinct from those summarized above. 
We establish the existence of a novel motility-driven concentration-wave instability that disrupts the orientationally ordered state even for $R>R_2$ (\cref{fig:phase-diag}). Following a linear stability analysis, we propose a minimal 1D model
to describe the physical mechanism of the novel instability. Finally, using high-resolution numerical simulations in two dimensions, we unveil three distinct states that arise from this new hydrodynamic instability: (1) Traveling waves, (2) a crossover regime featuring coexisting waves and defects, and (3) concentration-wave turbulence.

We obtain these results within a continuum description \cite{simha2002} of the active particle concentration $c({\bm x},t)$, the vector orientational order parameter $\pol({\bm x},t)$, and the hydrodynamic velocity ${\vel}({\bm x},t)$. The equations of motion \cite{simha2002,marchetti2013,giomi2012,cates2018,rana2024,kruse2005,chatterjee2021} read
\begin{align}
    \partial_t c + \bm{\nabla}\cdot \left[(\vel +v_{1}\pol)c\right] &= D\nabla^2 c \label{eq:conc}, \\
        \partial_t \pol+(\vel + v_0\pol)\cdot \bm{\nabla} \pol&=  \lambda \bm{S} \cdot \pol+ \Om \cdot \pol+ \Gamma \bm{h},~\mathrm{and} \label{eq:pol}, \\
    \rho (\partial_t  +  \vel \cdot \nabla) \vel&= -\bm{\nabla} P +\mu \nabla^{2}\vel+\bm{\nabla} \cdot \bm{\Sigma}, \label{eq:vel}
\end{align}
where $\rho$ is the constant suspension mass density and the pressure $P$ enforces the incompressibility constraint $\nabla\cdot\vel=0$. $v_0$ and $v_1$ are the speeds of self-advection and self-propulsion \cite{tjhung2011,cates2018,dadhichi2018, dadhichi2020, alert2022} respectively, the tensors $\Sij$ and $\Om$ are the symmetric and antisymmetric part of the velocity gradient tensor $\nabla\vel$, and $\lambda$ denotes the flow alignment parameter \cite{forster1974,larson1999}. The stress ${\bm \Sigma}={\bm \Sigma^r + \Sa}$ consists of the reversible thermodynamic  stress ${\bm \Sigma}^r \equiv -\lp \bm{h} \pol-\lm \pol\bm{h}$  with $\lambda_{\pm}\equiv (\lambda\pm 1)/2$ \cite{gennes1993, chandrasekhar1992, chatterjee2021}, and the apolar active stress $\Sa = -W c \pol\pol\equiv -\sa(c)\pol\pol$ \cite{ramaswamy2010, marchetti2013, alert2022} that models the swimmers as point force dipoles with $W>0$ for extensile swimmers \cite{daddi-moussa-ider2021}. The collective orientational mobility $\Gamma$ ($\sim 1/10\mu$ in molecular systems \cite{kneppe1983,gennes1993, mazenko1983, jadzyn2001}) governs the relaxational response of $\pol$ to the molecular field $\h\equiv -\delta F/\delta {\pol}$ with the free energy functional  
\begin{align} \label{eq:free-energy}
    F=\int\dd {\bm x} \left[-\frac{a(c)}{2}|\pol|^{2}+\frac{b}{4}|\pol|^{4}+\frac{K}{2}|\nabla \pol|^{2}+E \pol\cdot \nabla
    c\right].
\end{align}

For $a(c)<0$ and $b>0$, two homogeneous steady state solutions exist: the disordered state $\left({\vel}=0,c=0,|\pol|= 0\right)$, and an orientationally ordered state $\left({\vel}=0,c=c_0,|\pol|=p_0=\sqrt{|a(c_0)|/b}\right)$. We choose $a(c) = a \tanh m c$, with $a<0$ and $m>0$ which ensures that $p_0 \approx \sqrt{|a|/b}$ when $c \gg 1/m$  and $p_0 \to 0$ as $c \to 0$ \cite{giomi2012,chate2019}.
In \eqref{eq:free-energy}, a single Frank constant $K$ penalizes the spatial deformations in $\pol$ \cite{frank1958, oseen1933}. The parameter $E$ governs the alignment of $\pol$ to $\nabla c$. This term, analogous to the flexoelectric coupling in liquid crystals \cite{meyer1969}, is unique to polar systems because it breaks the $\pol \to -\pol$ symmetry \cite{kung2006}. In our system, as $\pol$ also governs the direction of self-propulsion, $E>0$ favors the movement of swimmers away from high-concentration regions whereas $E<0$, which we do not pursue here, would promote Motility-Induced Phase Separation (MIPS) \cite{cates2015,fily2012athermal,bialke2013,geyer2019}. A kinetic-theory derivation of the hydrodynamic equations \cite{bertin2006,bertin2009} for a dry system yields $\Gamma E c_0 \propto v_1$ but in general, the two are not related by any symmetry.

\emph{Linear Stability Analysis} -- The evolution of infinitesimal perturbations ${\bm \phi}\equiv(\delta c,\delta p_y, \delta u_x)$ to the homogeneous ordered steady state $(c=c_0,\pol=p_0\hat{x},\vel=\bm{0})$ is governed by the
linear system  
$\partial_t {\bm \phi}_{\mathbf q} (t) + {\mathsf M}({\mathbf q})\cdot {\bm \phi}_{\mathbf q}(t) = 0$, where $\bm{\phi} \equiv \bm{\phi}_{\bm{q}}\exp[i (\bm{q}\cdot{x} - \omega t)]$. 
Pure bend perturbations (${\bm q}=q \hat{x}$) do not couple to concentration fluctuations and their stability analysis is discussed in
\cite{chatterjee2021} and the Supplemental Material (SM) \footnote{see Supplemental Material, which includes Refs.~\cite{chatterjee2021, press2007, rana2024, bray2002, grassberger1983, mitra2018}. It contains details of the linear stability analysis, Pearson's correlation, the correlation dimension of the topological defects, and a description of the movies.}.
Crucially, for these perturbations, for $R$ larger than a threshold value $R_2\equiv \lp(1+\beta)^2/(2 \beta)$ with $\beta=\Gamma K \rho/\mu$, inertia stabilizes the ordered state (\cref{fig:phase-diag}).  

 For pure splay, i.e., for perturbations with $\bm{q} = q \hat{y}$, the dispersion matrix reads \cite{Note1}
\begin{equation}
{\mathsf M}({\bm q})=
\begin{pmatrix}
D q^2 &i v_1 c_0 q & 0 \\
i \Gamma E q & \Gamma  K   q^2 & -i p_0 \lm q\\
\frac{\lm p_0 E}{\rho} q^2 
&
\frac{ip_0 \left(W c_0 - K \lm q^2\right)}{\rho} q
& \frac{\mu}{\rho} q^2
\end{pmatrix}, 
 \label{eq:LS}
\end{equation}
leading to a novel instability which we describe below. 
For extensile ($W>0$), flow-tumbling ($|\lambda|<1$) swimmers \cite{giomi2008, giomi2012, amin2023}, a detailed examination of the dispersion relation reveals: (a) Splay-concentration waves \cite{toner1995,toner1998} modified by coupling to flow, propagating with speed $\sim \sqrt{c_0(\Gamma E v_1-W \lm p_0^2/\rho)}$ at $\mathcal{O}(q)$; and (b) a hitherto unremarked inertial instability at $\mathcal{O}(q^2)$ when
\begin{equation}\label{eq:threshold_E}
  \frac{\Gamma E c_0}{v_1} > \frac{f}{R},
\end{equation}
where $f$ depends on parameters in the equations of motion and is expected to be $\sim 0.1$ based on estimates for molecular or colloidal systems \footnote{For molecular systems, the rotational viscosity $\Gamma^{-1}   \sim 0.8$ poise \cite{kneppe1983,gennes1993}, the shear viscosity $\mu \sim 0.1$~poise \cite{mazenko1983,jadzyn2001}, 
the Frank constant $K\sim10^{-6}$~dyne \cite{gennes1993,bradshaw1985} and the diffusivity $D\sim 10^{-5}$ cm$^2$/s \cite{purcell1977, licata2016}. Therefore, for flow tumbling swimmers ($|\lambda| <1$) with $v_0=v_1$ and $p_0=1$, we get $\beta \sim 10^{-5}$ and $f\sim 0.1$. See Section I in SM for a detailed calculation of the stability criterion and the definition of $f$.}. 
The stability diagram in \cref{fig:phase-diag} highlights the canonical bend- and the new splay-unstable regimes. Note that either pure-splay or pure-bend are the dominant unstable modes depending on the value of $R$ and $\Gamma E c_0/v_1$. 

To expose the mechanism at the heart of the novel instability, hereafter we work with $\sigma_a = 0$ (zero active stress). In this limit, we obtain a simplified dispersion relation   
\begin{equation}\label{eq:dispersion_sigma0}
    \omega_\pm(q) = \pm q \sqrt{\Gamma E v_1 c_0} + \frac{i q^2}{2 \Gamma \rho}\left [{p_0^2 \lm^2} - \Gamma \rho(D+\Gamma K) \right], 
\end{equation}  
where $f$ in \eqref{eq:threshold_E} is inversely proportional to the square bracket on the right-hand side of \eqref{eq:dispersion_sigma0} \cite{Note1}. Note that $E$ is absent from the $\mathcal{O}(q^2)$ term in \eqref{eq:dispersion_sigma0}, but inequality \eqref{eq:threshold_E} ensures that the instability does not survive for $E=0$. Inertia, alignment of $\pol$ to $\nabla c$, and self-propulsion are all crucial for the novel instability. In the absence of any one of these couplings, the system \eqref{eq:LS} is linearly stable.

\emph{Instability mechanism} -- 
We now explain the 
mechanism of the $\mathcal{O}(q^2)$ instability within a minimal 1D model, retaining only the essential terms from the hydrodynamic equations that exhibit linear dispersion relations identical to \eqref{eq:dispersion_sigma0}. We restrict our analysis to gradients along
$\hat{y}$, that is, transverse to the ordering direction $\hat{x}$. The incompressibility condition then reduces to $\partial_y u_y = 0$, or $u_y$ is constant, which we set to zero.
For clarity, we denote $u_x = v$ and $p_y = p$ in this section. Consequently, the minimal 1D hydrodynamic equations for the variables $c(y,t)$, $p(y,t)$, and $v(y,t) $ are
\begin{align} \label{eq:1D}
    \begin{split}
        \partial_t c &= -v_1 \partial_y(pc) + D \partial^2_y c,\\
        \partial_t p &= \lm p_0 \partial_y v -\Gamma b p^3 + \Gamma K \partial^2_y p - \Gamma E \partial_y c,~\mathrm{and} \\
        \rho \partial_t v &= \mu \partial^2_y v + \lm p_0 E \partial^2_yc.
    \end{split}
\end{align}
We now show that inertia is essential for the novel instability. Consider first the Stokesian regime, where the balance of the viscous and reversible stresses instantly determines the velocity field as
\begin{equation} \label{eq:uxcstokes}
    \mu \partial^2_y v  \approx -\lm p_0 E \partial^2_yc.
\end{equation}
After eliminating $v$ from the $p$ equation we get
\begin{equation}
\partial_t p = -\left(\Gamma E + \lm^2 p_0^2 E/\mu\right) \partial_y c  -\Gamma b p^3 + \Gamma K \partial^2_y p, \end{equation}
where the concentration field couples to the order parameter via the pressure-like term $\partial_{y} c$, with an
additional contribution arising from the reversible stresses that accelerate the relaxation rate of the splay waves,
rendering the system \emph{stable}.

\begin{figure}
    \centering
    \includegraphics[width=\linewidth]{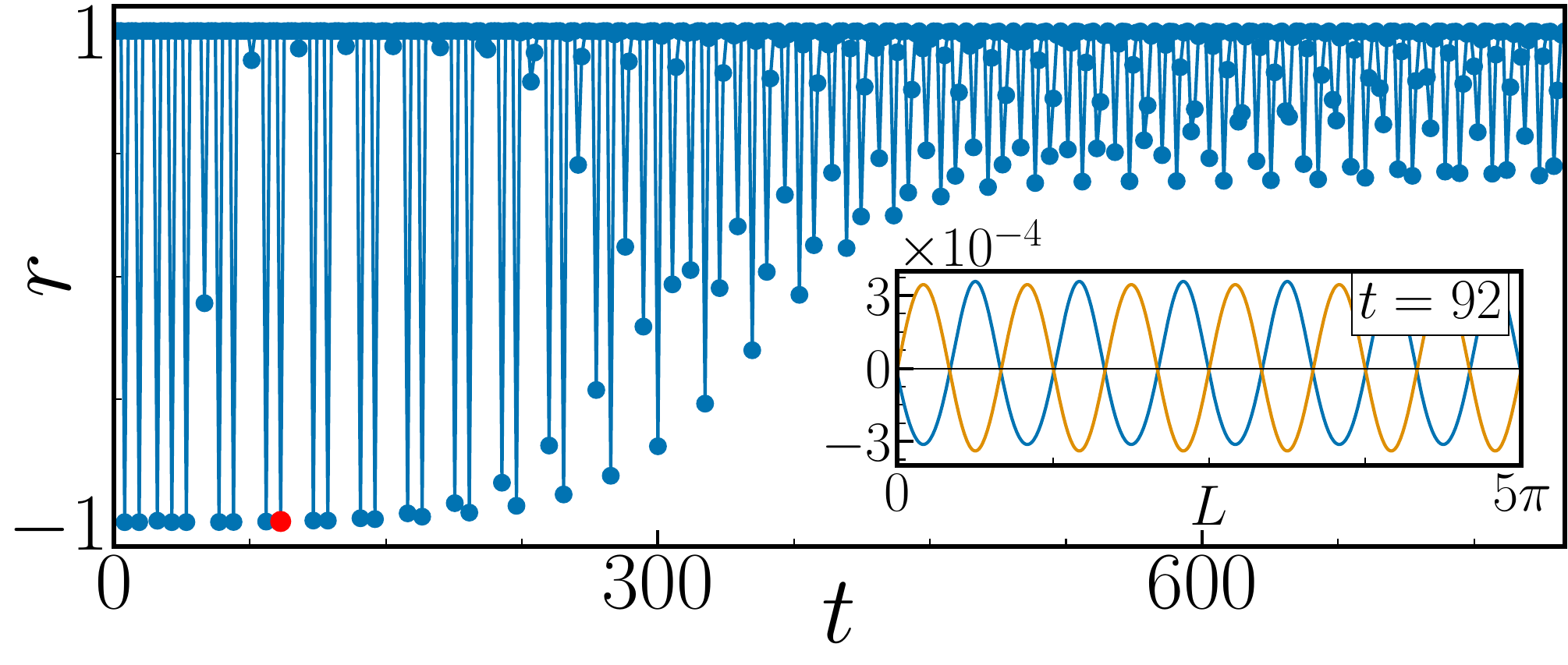}
    \caption{ \label{fig:corr}
        Pearson's correlation coefficient $r(t)$ at early times oscillates between $-1$ and $1$, implying varying temporal lag between $v$ and $c$. Inset: Velocity $v$ (orange) and concentration fluctuation $c - \overline{c}$ (blue) profiles at $t=92$ (marked by the red dot in the $r$ plot) during the onset of the instability. $v$ and $c$ are out-of-phase, which is crucial for the novel instability. 
    }
\end{figure}

Inertia introduces a crucial temporal lag between velocity and concentration that can negate the stability of the splay waves. While concentration attempts to relax, velocity may, at certain times, counteract this relaxation process. This out-of-phase interaction between velocity and concentration, driven by inertia, gives rise to the novel instability under consideration. To quantify our observations, we numerically integrate \eqref{eq:1D} for $E=0.2$ and plot the time evolution of Pearson's correlation coefficient $r(t)$ \cite{Note1} between the velocity and concentration fields in \cref{fig:corr}.

\begin{figure*}
    \centering
\includegraphics[width=0.9\linewidth]{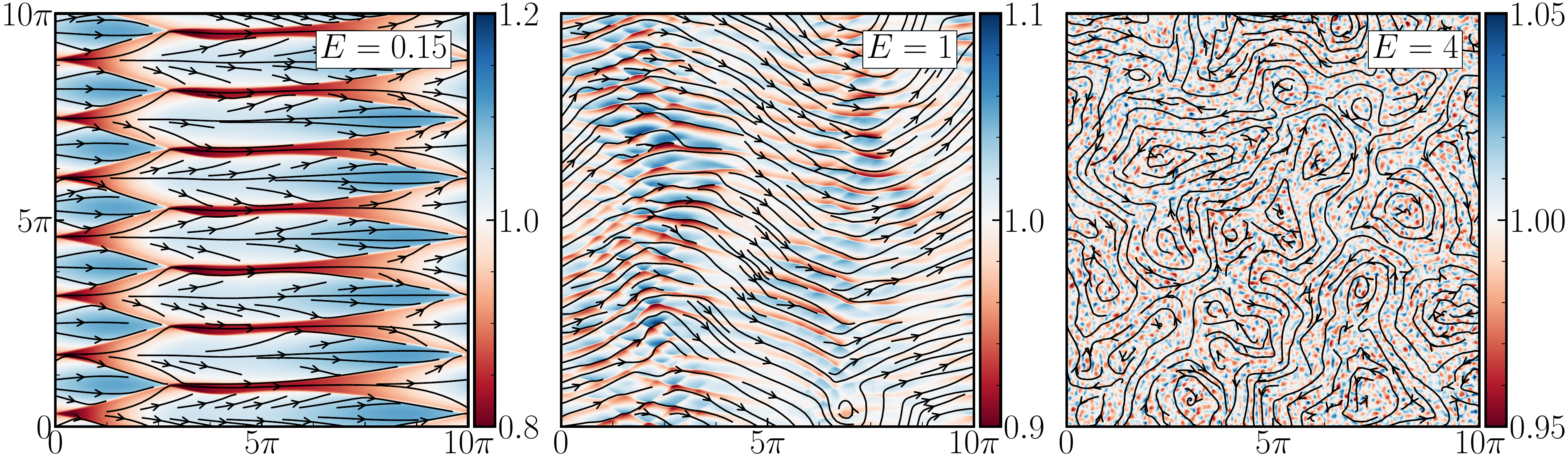}
    \caption{\label{fig:snapshots} 
    Pseudocolor plot of concentration with order parameter streamlines for varying $E$. As $E$ increases, disorder increases, and concentration fluctuations weaken. The system transitions from traveling waves at $E \leq 0.8$ to a crossover phase at 
    $0.8 < E < 4$, and to concentration-wave turbulence at 
    $E \gtrsim 4$.
     }
\end{figure*}

\begin{figure}
    \centering
    \includegraphics[width=\linewidth]{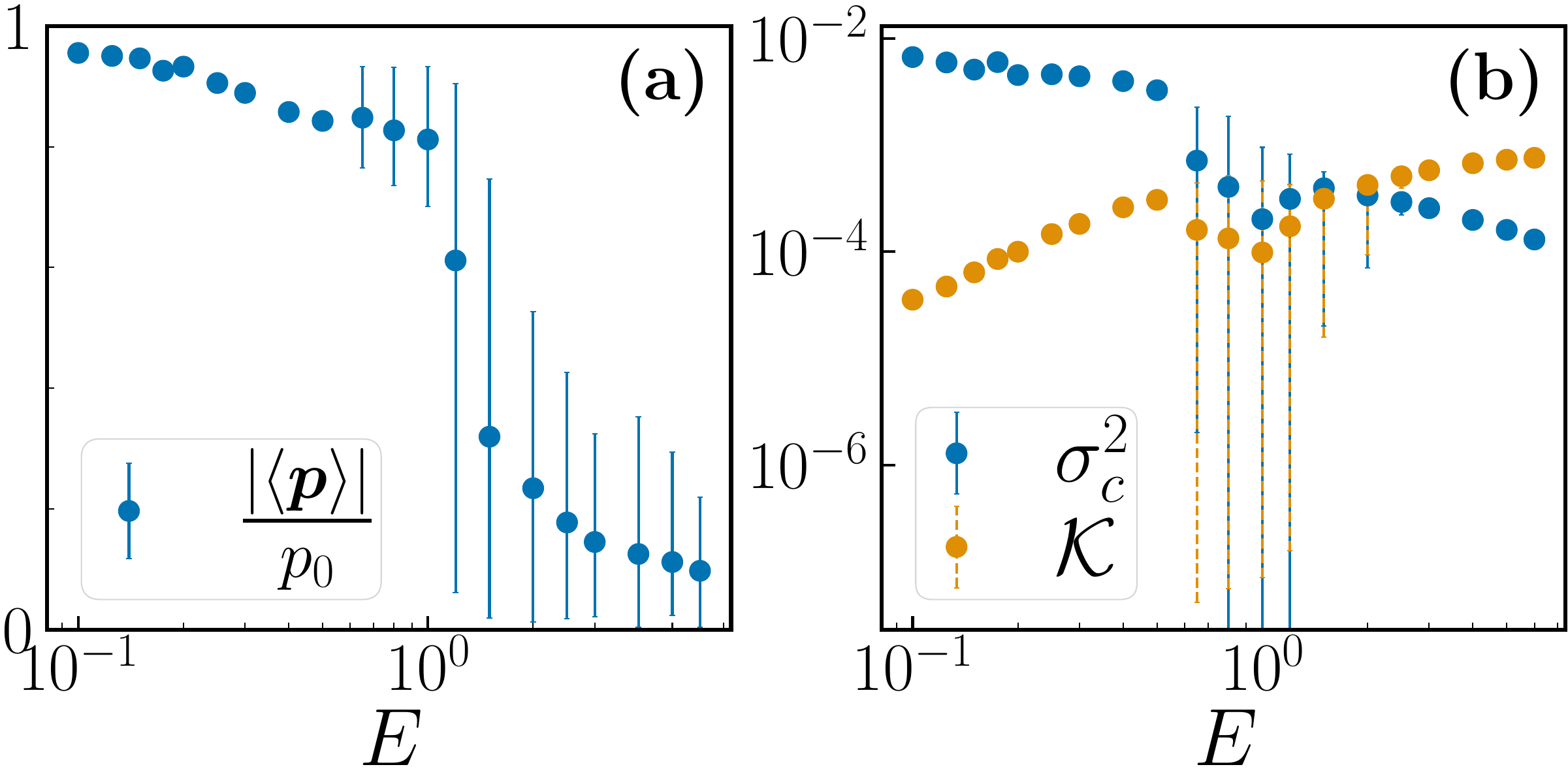}
    \caption{ \label{fig:var_avgp} Plot of the (a) average order $|\left<\pol\right>|$, (b) kinetic energy density $\KE$, and the concentration variance $\sigma_c^2$ versus $E$. With increasing $E$, $\sigma_c^2$ and $\avgp$ decrease and $\KE$ increases. Large fluctuations in the crossover phase are marked by error bars that correspond to the minima and maxima of the quantity.}
\end{figure}

\emph{Direct Numerical Simulations} -- To quantify the non-equilibrium steady states emerging from the novel instability, we numerically integrate \eqref{eq:conc}-\eqref{eq:vel} on a periodic square domain of length $L=10\pi$ discretized with $N^2=1024^2$ points. As mentioned earlier, we set $\sa=0$ to emphasize the effects of the novel instability. We use the streamfunction-vorticity formulation for the numerical integration of \eqref{eq:vel} \cite{perlekar2009}. The spatial derivatives are evaluated using a fourth-order centered finite-difference scheme, and a second-order Adams-Bashforth scheme is used for time integration.
In what follows, we study the statistically steady states with varying $E$ while keeping other parameters fixed ($\rho=1$, $\mu=0.1$, $W=0$, $\Gamma = 1$, $D=10^{-4}$, $K=10^{-3}$, $\lambda=0.1$, $v_1=v_0=0.1$, $a=-0.1$, $m=10$, $b=1$, and $c_0=1$). To be consistent with molecular systems \cite{Note2}, we choose parameters such that $\mu \gg \rho \Gamma K, \rho D$,  and $\rho K/\mu^2 \ll 1$. This amounts merely to assuming that velocity gradients relax much faster than that of orientation and concentration.
\cref{fig:snapshots} shows the snapshots of the concentration field and streamlines of the order parameter field in the steady state for various values of $E$. As $E$ increases, disorder sets in, and fluctuations in the concentration field decrease. We quantify the same in \cref{fig:var_avgp}, which shows the plot of the average order $|\langle \pol \rangle|$, the variance in concentration $\sigma_c^2\equiv \langle c^2\rangle -  \langle c \rangle^2$, and the kinetic energy density $\KE \equiv \langle {\vel}^2/2\rangle$. Here, the angular brackets denote spatiotemporal averaging in the steady state. Consistent with the snapshots, $|\langle \pol \rangle|$ decreases as $E$ increases. For $E \leq 0.8$ and $E > 2$, $\KE$ increases monotonically, whereas $\sigma_c^2$ decreases. In between, we find a short cross-over region where $|\langle \pol \rangle|$, $\sigma_c^2$, and $\KE$ all exhibit large fluctuations as marked by significant error bars in \cref{fig:var_avgp}. A careful inspection of \cref{fig:snapshots,fig:var_avgp}, and other quantifiers shows the presence of three distinct regimes we now describe.

\begin{figure}
    \centering
    \includegraphics[width=0.9\linewidth]{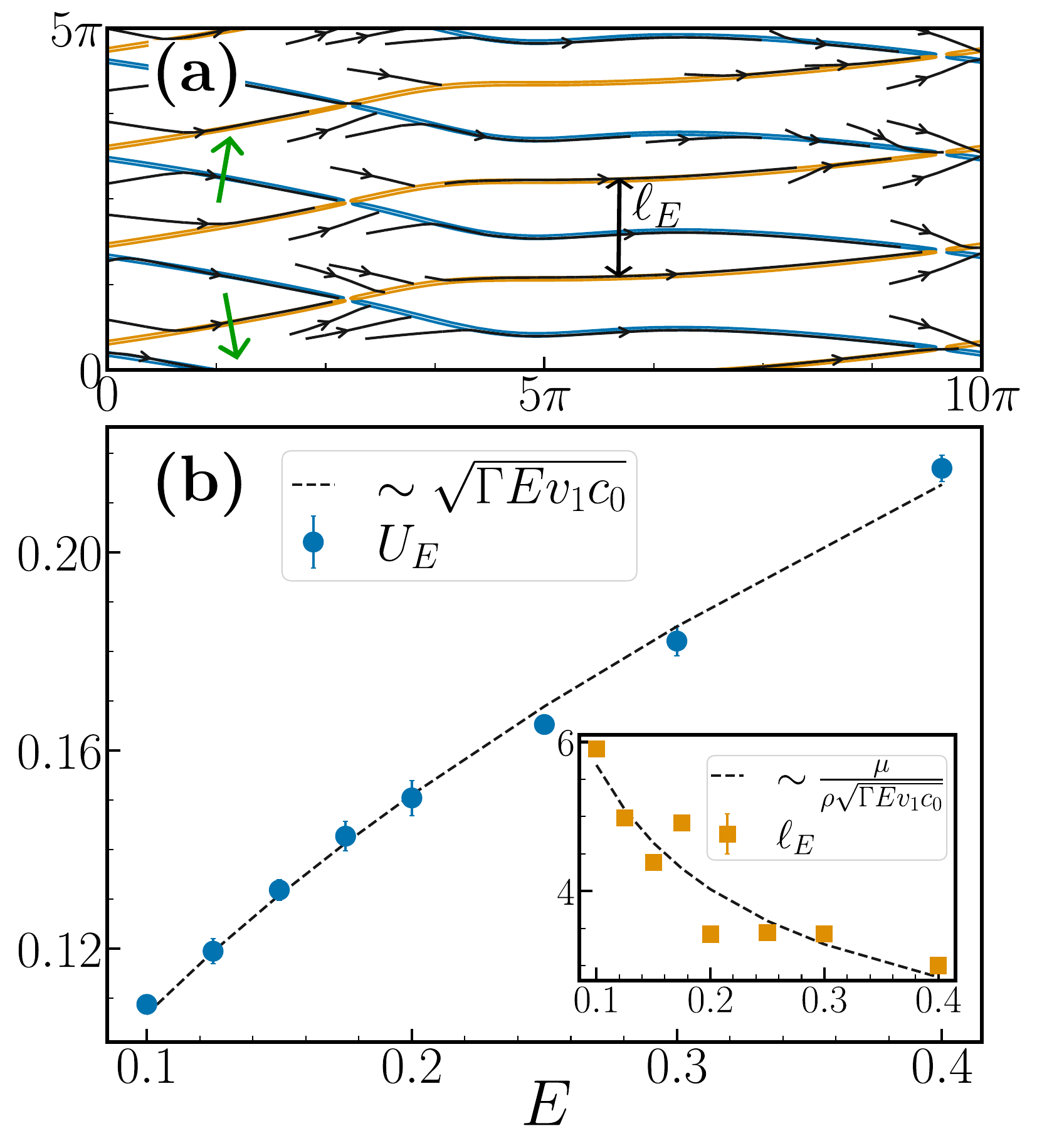}
    \caption{ \label{fig:wavespeed} (a) Iso-contours of $|\bdel c|=0.5$ with streamlines of $\pol$ (black lines) for $E=0.15$. Contours with $\bdel c \cdot \hat{y} >0 (<0)$ are drawn with orange (blue) lines. Solid Green arrows show the direction in which the wavefront travels. (b) Wave speed $U_E$ and (Inset) distance between wavefronts $\ell_E$ versus $E$. (Black dashed line) Prediction based on dimensional analysis  $U_E \sim \sqrt{\Gamma E v_1 c_0}$ and $\ell_E \sim \mu/{(\rho \sqrt{\Gamma E v_1 c_0})}$.}

\end{figure}

I. \emph{Traveling waves --}
For  $E \le 0.8$, we observe traveling waves. \cref{fig:snapshots} ($E=0.15$) shows a typical realization of the concentration field in this regime. In \cref{fig:wavespeed}(a), we plot $|\bdel c|$ for $E=0.15$
and identify with iso $|\bdel c|=0.5$ contours, two wavefronts moving opposite each other in the direction  $-\bdel c/|\bdel c|$.
Intriguingly, the wavefronts behave like solitons, i.e., they move with a constant speed and preserve their shape as they pass through each other. We track the wavefronts to evaluate their speed $U_E$ and the wavelength $\lE$ plotted in \cref{fig:wavespeed}(b). Consistent with the dimensional analysis, we find $U_E\sim \sqrt{\Gamma E v_1 c_0}$ and $\lE\sim \mu/(\rho \sqrt{\Gamma E v_1 c_0})$.

II. \emph{The crossover phase --} For $0.8<E<4$, waves and defects coexist. The wave trains discussed in the previous section are destabilized, and we observe the spontaneous appearance of vortical structures in the order parameter field (See $E=1$ in \cref{fig:snapshots}). Consistent with the large error bars in
$\sigma_c^2$ in \cref{fig:var_avgp}(b), the plot of $\overline{c^2}(t) = (1/L) \int c^2 d{\bm x}$ in the steady state shows large fluctuations. Small values of $\overline{c^2(t)}-\overline{c(t)}^2 < 10^{-4}$ correspond to the nearly homogeneous $c$ and nearly ordered $\pol$ field \cite{Note1}.

III. \emph{Concentration-wave turbulence $(E>4)$ ---} To resolve large and small-scale structures in the turbulent regime, we perform large-scale simulations at $L=100\pi$ with a grid resolution of $N^2=8192^2$ while keeping other parameters fixed to the same value. The time evolution of the concentration field shows the presence of traveling concentration waves, and the order parameter streamlines show complex spatiotemporal structures [See \cref{fig:snapshots} ($E=4$) and the movie in SM \cite{Note1}]. These concentration waves resemble that observed in Regime I and their wavelength $\lE \sim \mu/(\rho\sqrt{\Gamma E v_1 c_0})$.

\begin{figure}
\centering
\includegraphics[width=0.9\linewidth]{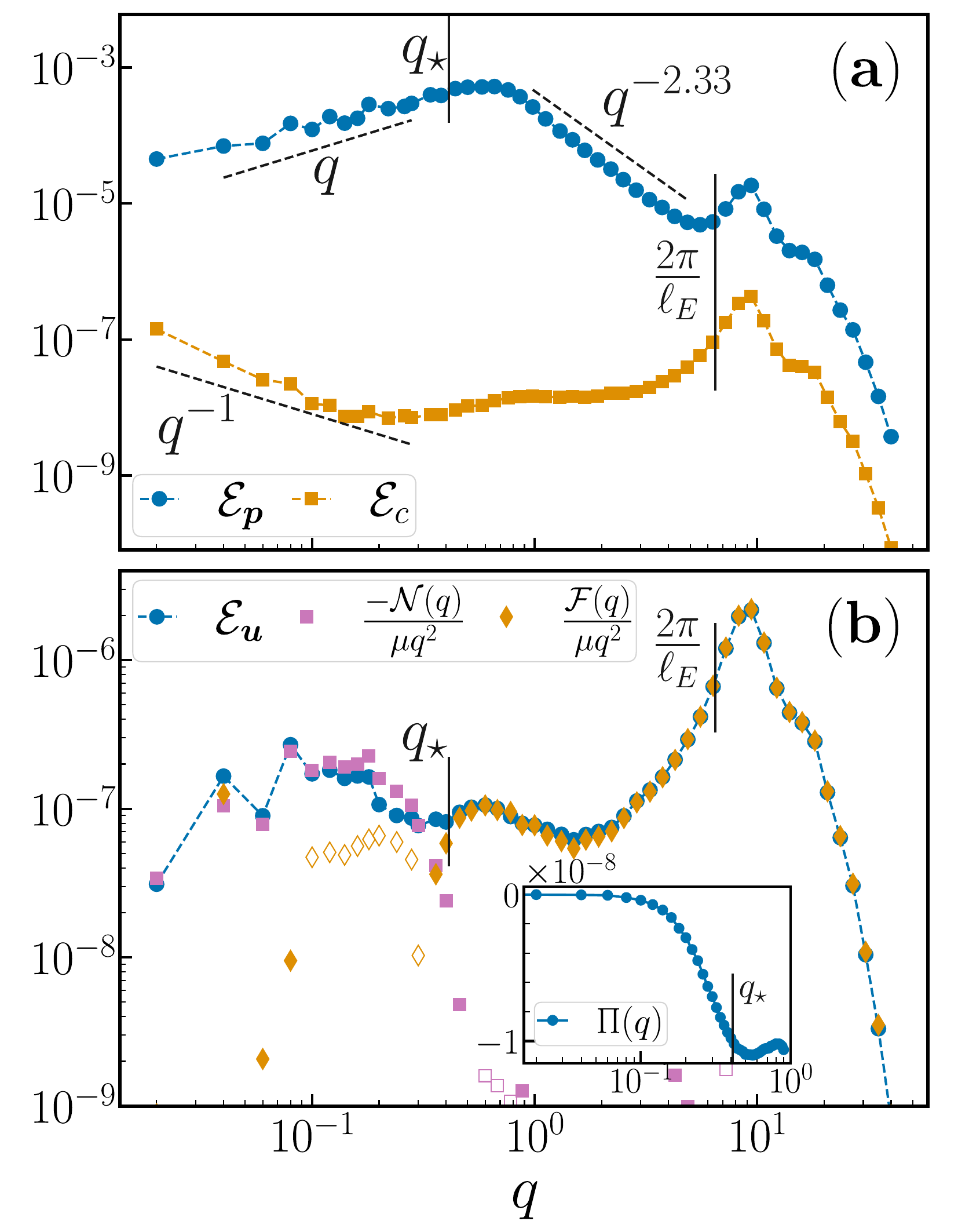}
    \caption{\label{fig:spectra} Plot of the (a) order parameter spectra $\Ep$, the concentration spectra $\Ec$, and (b) the kinetic energy spectra $\Eu$ for $E=4$. All the spectra show a peak around $2\pi/\lE$ where $\lE \sim 0.97$. (a) $\Ep$ peaks at  $q_\star \sim 0.41$ and shows a power law scaling $\sim q^{-(1+d_2)}$ for $q_\star < q < 2\pi/\ell_E$, where $d_2=1.33$ is the correlation dimension of the topological defects. $\Ec$ shows Batchelor scaling for $q<q_\star$. (b) The dominant balance \eqref{eq:balance} matches $\Eu$ well. Open markers indicate the negative value of the corresponding quantity. (Inset) Energy flux $\Pi(q)$ versus $q$.
   } 
\end{figure}
We characterize the turbulent regime using the power spectrum of the order parameter, concentration, and velocity fields i.e.,
\begin{align}
\Ep \equiv \sum_{\bm k}^\prime |\hat{\bm \pol}_{\bm k}|^2,~
\Ec \equiv \sum_{\bm k}^\prime |\hat{c}_{\bm k}|^2,~\mathrm{and}~
\Eu \equiv \sum_{\bm k}^\prime |\hat{\vel}_{\bm k}|^2,
\end{align}
where $\hat{()}_{\bm k}$ denotes the Fourier coefficient of the wavenumber $\bm k$ and $\sum_{\bm k}^\prime$ implies a sum over all Fourier modes that satisfy $q - \pi/L \leq |\bm k| < q + \pi/L$.
In \cref{fig:spectra} we plot the time-averaged $\Ep,~\Ec,$ and $\Eu$ in the steady state and observe a peak around $q\sim 2\pi/\lE$ signaling the dominant wavenumber of the concentration waves. $\Ep$ shows a small-$q$ peak at $q \sim q_\star$, which is inversely proportional to the typical large-scale inter-vortex separation.
$\Ep$ has two power law regimes around $q_\star$. For $q\ll q_\star$, we obtain equipartition spectrum $\Ep \sim q$, indicating that the fluctuations in the order parameter field are uncorrelated. In the regime $q_\star < q < 2\pi/\lE$, we find a modified Porod's scaling $\Ep\sim q^{-(1+d_2)}$, where $d_2$ is the correlation dimension which characterizes the clustering of the topological defects over these length scales \cite{rana2024, Note1}. For reasons that are unclear at this stage, the profile of the spectrum and the exponents are close to those seen in models of bacterial turbulence, where activity is modeled as energy injection in a band of wavenumber \cite{wensink2012}.  

At large length scales ($q\ll q_\star$), $\Ep \gg \Eu$, thus the concentration field is primarily advected by random fluctuations of the order parameter. As the Schmidt number $Sc\equiv \mu/(\rho D) \gg 1$, we observe Batchelor scaling $\Ec \sim q^{-1}$ for $q<q_\star$, establishing a nice connection to the phenomenology of stochastically advected passive scalars \cite{batchelor1959,kraichnan1974}.

Finally, to understand the scaling of the kinetic energy spectrum, we derive the following dominant steady-state energy balance \cite{frisch1995}:
\begin{align}\label{eq:balance}
   {\mathcal{N}}(q) \approx -\mu q^2  \Eu  +  {\mathcal{F}}(q),
\end{align}
where  the nonlinear transfer ${\mathcal{N}}(q) = \rho \sum_{\bm k}^\prime \mathrm{Re}[\hat{\vel}(-{\bm k}) \cdot (\mathcal{P}\cdot\widehat{[\vel \cdot \nabla \vel]}({\bm k})]$, the contribution due to the aligning stresses ${\mathcal{F}}(q) = \sum_{\bm k}^\prime \mathrm{Re}[\hat{\vel}(-\bm k) \cdot (\mathcal P \cdot (i \bm{k}\cdot\hat{\bm \Sigma}^E({\bm k}))]$ with $\bm{\Sigma}^E_{ij} = E(\lp p_j\partial_i c + \lm p_i \partial_j c)$, and $\mathcal{P}={\mathcal I}- {\bm q}{\bm q}/q^2$ is the projection operator.

For $q>q_\star$, viscous dissipation balances the energy injected by aligning stress $\bm{\Sigma}^E$. In contrast, for $q<q_\star$, we observe negative energy flux $\Pi(q)=\sum_{k<q} \mathcal{N}(k)<0$ (\cref{fig:spectra}(b,inset)),
which implies an inverse energy transfer to large scales \cite{boffetta2012, pandit2009, perlekar2017,koch2021,rorai2022} balanced primarily by viscous dissipation. Recent works on active nematics \cite{koch2021,rorai2022} also show that the presence of inertia can lead to an inverse energy cascade similar to hydrodynamic turbulence.

\emph{Conclusions ---} We identify a hitherto unknown inertial instability in an active suspension of polar particles, which does not rely on the widely studied active stress \cite{simha2002,chatterjee2021,rana2024,giomi2012,marchetti2013}.
Through extensive numerical simulations, we examine how varying the coupling strength between the polar order parameter and the concentration field affects the statistical steady states, identifying three regimes: I) Traveling waves, II) Crossover phase, and III) Concentration-wave turbulence. In the discussions above we had chosen $W=0$. However, as indicated by the stability diagram (\cref{fig:phase-diag}), we confirm numerically that the turbulence discussed above is also observed for small $R=0.1$ or finite $W=0.05$ but with a large value of $\Gamma E c_0/v_1=40$ \cite{Note1}. Active systems with intermediate $\Rey$ at the scale of the motile components, such as concentrated suspensions of zooplankton ($\Rey \sim 1 - 100$) \cite{Wadhwa2015,kiorboe2014, anderson2015, Tarling2016}, are candidate settings in which to seek the novel concentration-wave instability and turbulence that we predict. Such studies, however, would be exploratory. A systematic test of our prediction \eqref{eq:threshold_E} could be carried out by varying the swimming speed in a controllable model suspension of motile particles at moderate Reynolds number, with a spontaneously aligned and moving phase. We hope our work will motivate such experiments in this largely unexplored domain of inertial active fluids.

PJ and PP  acknowledge support from the Department of Atomic Energy (DAE), India under Project Identification No. RTI 4007, and DST (India) Project Nos. MTR/2022/000867,  and DST/NSM/R\&D HPC Applications/Extension/2023/08. 
SR gratefully acknowledges a J C Bose Fellowship of the SERB, India. SR and PP thank the Isaac Newton Institute for Mathematical Sciences, Cambridge, for support and hospitality during the programme \href{https://www.newton.ac.uk/event/adi/}{Anti-diffusive dynamics: from sub-cellular to astrophysical scales} where work on this paper was undertaken, supported by EPSRC grant no EP/R014604/1. All the simulations are performed using the HPC facility at TIFR Hyderabad.

\end{document}


\title{Supplementary Material for Inertia drives concentration-wave turbulence in swimmer suspensions}

\author{Purnima Jain}
\affiliation{Tata Institute of Fundamental Research, Hyderabad, India}
\author{Navdeep Rana}
\affiliation{Max Planck Institute for Dynamics and Self-Organization (MPIDS), D-37077 G\"ottingen, Germany}
\author{Sriram Ramaswamy}
\affiliation{Indian Institute of Science, Bangalore, India}
\author{Prasad Perlekar}
\email{perlekar@tifrh.res.in}
\affiliation{Tata Institute of Fundamental Research, Hyderabad, India}

\maketitle

\tableofcontents

\setcounter{equation}{0}
\setcounter{figure}{0}
\renewcommand{\theequation}{S\arabic{equation}}
\renewcommand{\thefigure}{S\arabic{figure}}

\section{Linear stability Analysis with active stress}
We study the linear stability of the following equations about the homogeneous ordered steady state ($\vel=0,~|\pol| = p_0= \sqrt{|a(c_0)|/b},~c=c_0$),
\begin{align}
 \partial_t c + \bm{\nabla}\cdot \left[(\vel +v_{1}\pol)c\right] &= D\nabla^2 c, \\
 \partial_t \pol +(\vel + v_0\pol)\cdot \bm{\nabla} \pol &=  \lambda \bm{S} \cdot \pol + \bm{\Omega} \cdot \pol + \Gamma \bm{h},\mathrm{and} \\
    \rho (\partial_t  + \vel \cdot \bm{\nabla})\vel &= - \bm{\nabla} P + \mu \nabla^{2}\vel + \bm{\nabla} \cdot (\Sa + \bm{\Sigma^r}).
\end{align}
Here,
    ${\bm \Sigma}^r \equiv -\lp \bm{h} \pol -\lm \pol \bm{h}$  with $\lambda_{\pm}\equiv (\lambda \pm 1)/2$, $\Sa = -W c \pol \pol \equiv -\sa(c) \pol \pol$, $\Sij \equiv \frac{\bm \nabla \vel + \bm \nabla \vel^T}{2}$, $\Om \equiv \frac{\bm \nabla \vel - \bm \nabla \vel^T}{2}$, and $\bm h = (-a(c)+b|\pol|^2)\pol + K \nabla^2 \pol - E \bm \nabla c$ with $E>0$. We choose monochromatic perturbations of the form
\begin{align}
{\bm \phi_{\bm q}(t) \equiv}
\begin{pmatrix} \delta c \\ \delta p_y \\ \delta u_x \end{pmatrix}
=
\begin{pmatrix}\hat{c} \\ \hat{p} \\ \hat{u} \end{pmatrix}
e^{i (\mathbf{q}\cdot \bm{r} - \omega t)},
\end{align}
in the direction transverse to the ordering direction that we take to be $\hat{x}$, therefore $\delta p_x=0$.
The system is linearly unstable where $\mathrm{Im}(\omega)>0$ and stable otherwise. 
The linearised system obtained from the hydrodynamic equations is,
\begin{align}
 \partial_t {\bm \phi}_{\mathbf q} (t) + {\mathsf M}({\mathbf q})\cdot {\bm \phi}_{\mathbf q}(t) = 0,
\end{align}
where
\begin{equation}
{\mathsf M}({\mathbf q})=\begin{pmatrix}
i v_1 p_0 q_x + D q^2 & i v_1 c_0 q_y & 0 \\
i \Gamma E q_y & i v_0 p_0 q_x +\Gamma  K q^2 & \frac{i p_0}{q_y} (\lp q_x^2 - \lm q_y^2)\\
q_y^2\left(\frac{i p_0^2 q_x}{q^2} \left(W +\lambda a'(c_0)\right) + \lm p_0 E\right)/\rho
&
i p_0 q_y\left(K\left(\lp q^2 - \lambda q_y^2\right) - W c_0 \frac{(q_x^2 - q_y^2)}{q^2}\right)/\rho
&  \mu q^2/\rho
\end{pmatrix}.
\label{eq:LSA}
\end{equation}

\subsection*{Pure bend instability}
For pure bend modes ($\bm q = q \hat{x}$), the incompressibility of the velocity field ($q_x \delta u_x + q_y \delta u_y = 0$) yields $\delta u_x = 0$. Therefore, the dispersion matrix \eqref{eq:LSA} is modified by substituting $\delta u_x = -q_y \delta u_y/q_x$, and
\begin{align}
{\bm \phi \equiv}
\begin{pmatrix} \delta c \\ \delta p_y \\ \delta u_y \end{pmatrix}
=
\begin{pmatrix}\hat{c} \\ \hat{p} \\ \hat{u} \end{pmatrix}
e^{i (\mathbf{q}\cdot \bm{r} - \omega t)}.
\end{align}
The dispersion matrix obtained from \eqref{eq:LSA} is,
\begin{equation}
{\mathsf M}({\mathbf q})=\begin{pmatrix}
i v_1 p_0 q + D q^2 & 0 & 0 \\
0 & i v_0 p_0 q +\Gamma K q^2 & -i p_0 \lp q\\
0
&
i p_0 q (W c_0 - \lp K q^2)/\rho
&  \mu q^2/\rho
\end{pmatrix}.
\label{eq:LSA_pb}
\end{equation}
The pure bend modes decouple concentration from the order parameter and velocity fields yielding two eigenmodes. In the hydrodynamic limit ($q \to 0$), up to $\mathcal{O}(q^2)$, we obtain
\begin{equation}
    \omega_\pm (q)= \frac{v_0 p_0 q}{2} \pm \frac{v_0 p_0 q}{2}\sqrt{1-\frac{(1+\lambda)}{R}} - \frac{i q^2 \mu}{2\rho}\left(1+\beta \mp \frac{(1-\beta)}{\sqrt{1-\frac{(1+\lambda)}{R}}} \right).
\end{equation}
The dispersion relation obtained is same as in \cite{chatterjee2021, rana2024} and it yields a bend instability at $\mathcal{O}(q)$ for $R<R_1 = (1+\lambda)$ and $\mathcal{O}(q^2)$ for $R_1<R<R_2$ where $R_2 = \lp(1+\beta)^2/2\beta$, $\beta=\Gamma K\rho/\mu$, and $R=\rho v_0^2/(2 W c_0)=\rho v_0^2/2\sa(c_0)$.

\subsection*{Pure splay instability}
Pure splay modes ($\bm q = q \hat{y}$) couple concentration fluctuations with the order parameter and velocity fields yielding three eigenmodes ($\omega_{1,2,3}$). The following linear dispersion matrix obtained from \eqref{eq:LSA} is
\begin{equation}
{\mathsf M}({\mathbf q})=\begin{pmatrix}
D q^2 &i v_1 c_0 q & 0 \\
i \Gamma E q & \Gamma  K   q^2 & -i p_0 \lm q\\
\lm p_0 E q^2/\rho
&
\left(-i p_0 K \lm q^3 + i W c_0 p_0 q\right)/\rho
& \mu q^2/\rho
\end{pmatrix}.
\end{equation}
In the hydrodynamic limit $q \to 0$, up to $\mathcal{O}(q^2)$, we get, 
\begin{align}
 \omega_1 &= - i q^2\left(\frac{-W D \lm p_0^2 + \Gamma E v_1 \mu + E v_1 \lm^2 p_0^2}{\Gamma E v_1 \rho - W \lm p_0^2 }\right) + \mathcal{O}(q^3), \mathrm{and}\\
 \omega_{2,3} &= \pm q \sqrt{\left(\Gamma E v_1-\frac{W \lm p_0^2}{\rho}  \right)c_0} + \frac{i q^2}{2 \rho}\left(\frac{E v_1 \rho \left(\lm^2 p_0^2-\Gamma \rho (D+\Gamma K) \right) + W \lm p_0^2(\mu + \Gamma K \rho)}{\Gamma E v_1 \rho - W \lm p_0^2 }\right) + \mathcal{O}(q^3). \label{eq:LSA_ps}
\end{align}
For extensile ($W >0$), flow-tumbling ($|\lambda|<1$) or flow-aligining oblate ($\lambda<-1$) swimmers, we get one stable ($\omega_1$) and two unstable modes ($\omega_{2,3}$) with equal growth rates. The novel concentration-wave instability at $\mathcal{O}(q^2)$ arises if
\begin{equation}\label{eq:threshold_E}
  \frac{\Gamma E c_0}{v_1} > \frac{f}{R},
\end{equation}
with 
\begin{equation}
f=\left(\frac{v_0^2}{2 v_1^2}\right)\left(\frac{|\lm| p_0^2\Gamma \mu(1 + \beta)}{\lm^2 p_0^2-\Gamma \rho (D+\Gamma K)}\right).
\end{equation}

In contrast, flow-aligning prolate  swimmers ($\lambda >1$) are splay unstable at $\mathcal{O}(q)$ if $W \lm p_0^2/\rho > \Gamma E v_1$ \eqref{eq:LSA_ps}. However, this instability is not the focus of present work and will be addressed in future studies.

\REM{Note that for molecular systems, the rotational viscosity $\Gamma^{-1}   \sim 0.8$ poise \cite{kneppe1983,gennes1993}, the shear viscosity $\mu \sim 0.1$~poise \cite{mazenko1983,jadzyn2001}, 
the Frank constant $K\sim10^{-6}$~dyne \cite{gennes1993,bradshaw1985} and the diffusivity $D\sim 10^{-5}$ cm$^2$/s \cite{purcell1977, licata2016}. Therefore, for flow tumbling swimmers ($|\lambda| <1$) with $v_0=v_1$ and $p_0=1$, we get $\beta \sim 10^{-5}$ and $f\sim 0.1$.}

\begin{figure}[h!]
 \includegraphics[width=0.8\linewidth]{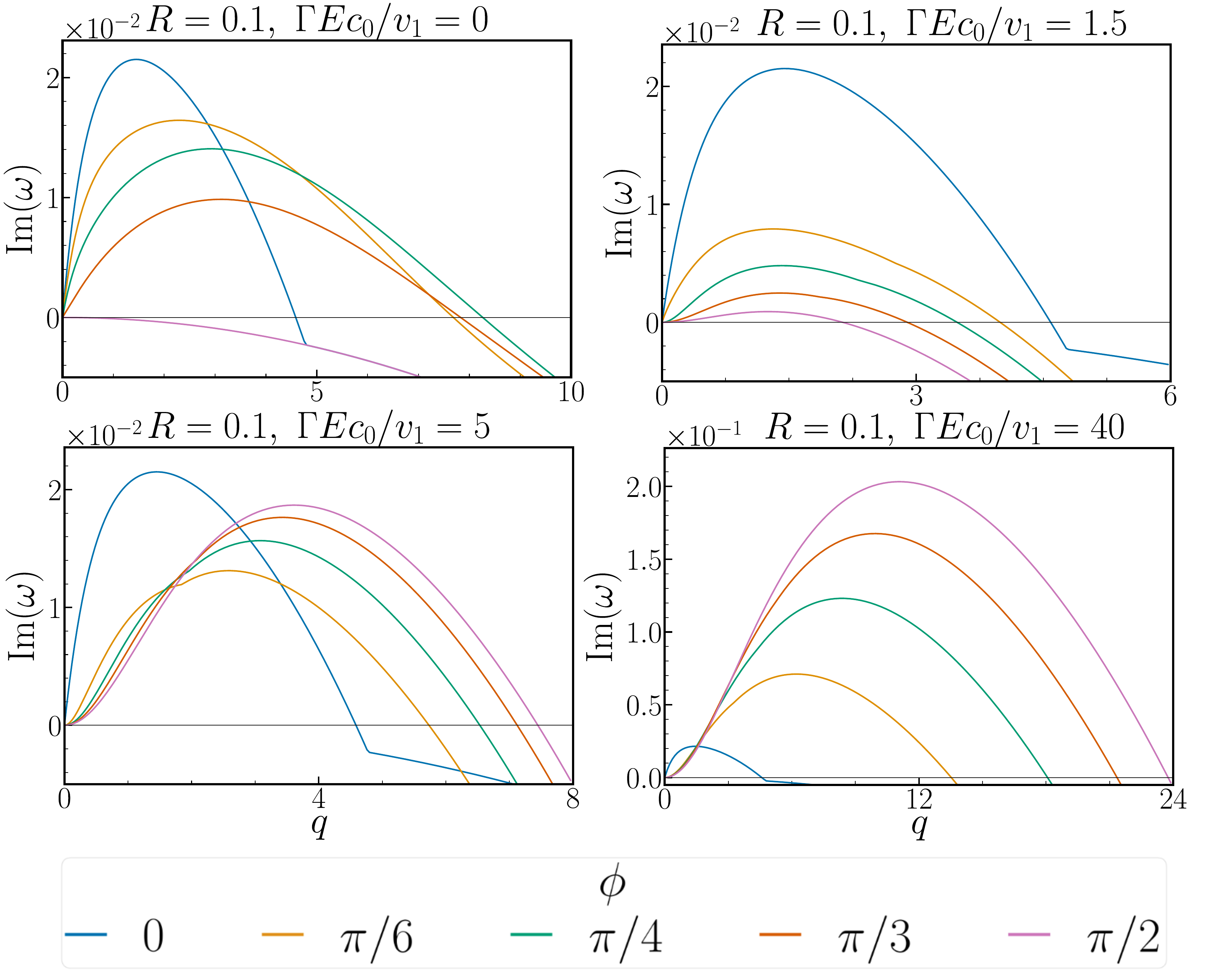}
 \caption{ \label{fig:dispersion_R0.1}
Dispersion curves for different values of $\Gamma E c_0/v_1$ for $R=0.1~(W=0.05)$. For $\Gamma E c_0/v_1=0$, the system is splay stable with the bend instability being dominant. As $\Gamma E c_0/v_1$ increases beyond the threshold value, the new splay-concentration instability appears and becomes dominant for a large value of $\Gamma E c_0/v_1$. The other parameters are: $\rho=1$, $\mu=0.1$, $\Gamma=1$, $D=10^{-4}$, $K=10^{-3}$, $\lambda=0.1$, $v_1=v_0=0.1$, $p_0=\sqrt{0.1}$, and $c_0=1$.}
\end{figure}

The above analysis shows that in the presence of inertia and concentration fluctuations, a novel splay instability arises for extensile swimmers that were otherwise stable for large values of $R>R_2$ \cite{chatterjee2021,rana2024}. This instability is present in an inertial system along with the bend instability, with either of them dominating depending on the value of $R$ and $\Gamma E c_0/v_1$. \cref{fig:dispersion_R0.1} shows the dispersion curves for $R=0.1~(W=0.05)$ with varying $\Gamma E c_0/v_1$ and angle of perturbation ($\phi$). For the parameters used, $R_2 \approx 28.05$ and threshold $\Gamma E c_0/v_1 \approx 1.2$. For $\Gamma E c_0/v_1=0$, the system is splay stable and the bend instability is dominant. Above the threshold $\Gamma E c_0/v_1$, the novel splay instability appears and becomes dominant for large values of $\Gamma E c_0/v_1$.

\section{Linear Stability Analysis without active stress}
In the following, we ignore active stress ($\Sa=0$) and perform the linear stability analysis in a similar way as above. The linearised system obtained from \eqref{eq:LSA} is, 
\begin{equation}
\mathsf{M} (\bm{q}) = 
\begin{pmatrix}
i v_1 p_0 q_x + D q^2 &i v_1 c_0 q_y & 0 \\[8pt]
i \Gamma E q_y & i v_0 p_0 q_x +\Gamma  K   q^2 &\frac{i p_0}{q_y} (\lp q_x^2 - \lm q_y^2)\\[8pt]
\left( \lm E p_0   q_y^2 + \frac{i \lambda a'(c_0) p_0^2 q_y^2 q_x }{q^2} \right)/\rho
&
i p_0 K q_y(\lp q^2 - \lambda q_y^2)/\rho
&  \mu q^2/ \rho
\end{pmatrix}.
\label{eq:LSA_sigma0}
\end{equation}
The system is linearly stable for pure bend modes ($\bm q=q \hat{x}$) that decouple concentration from velocity and order parameter fields. For pure splay modes ($\bm q=q \hat{y}$), we get, 
\begin{equation}
{\mathsf M}({\mathbf q})=
\begin{pmatrix}
 D q^2& i v_1 c_0 q  & 0 \\
i \Gamma E q &  \Gamma K q^2 & -i p_0 \lambda_- q \\
\lm E p_0 q^2/\rho & - i p_0 K \lm q^3/\rho  & \mu q^2/\rho 
\end{pmatrix}.
\label{eq:LSA_pb}
\end{equation}

\begin{figure}
 \includegraphics[width=\linewidth]{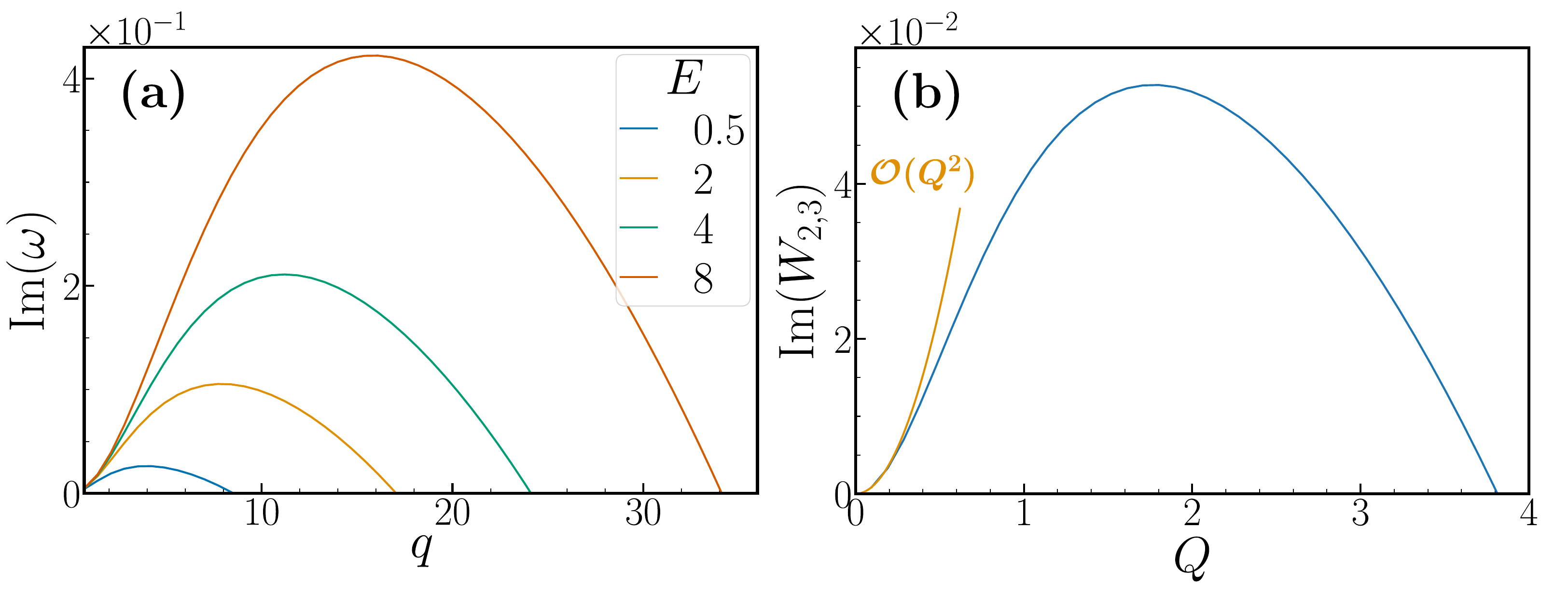}
 \caption{ \label{fig:dispersion}
  (a) Dispersion curves for different $E$ for pure splay modes. $\rho=1$, $\mu=0.1$, $W=0$, 
 $\Gamma = 1$, $D=10^{-4}$, $K=10^{-3}$, $\lambda=0.1$, $v_1=v_0=0.1$, $p_0=\sqrt{0.1}$, and $c_0=1$. (b) Dimensionless dispersion relation Im($W_{2,3}$) versus $Q$ for pure splay.}
\end{figure}

The characteristic equation for the above matrix is a cubic polynomial in $\omega$,

\begin{equation}
    \begin{split}
    i \omega^3 - \omega^2 q^2 \left(D+\Gamma K + \frac{\mu}{\rho}\right) -i \omega \left[\Gamma E v_1 c_0 q^2 + q^4\left(\frac{\lm^2 p_0^2 K + \mu D + \Gamma K \mu + \Gamma K D \rho}{\rho}\right) \right] \\
    + \frac{\Gamma E v_1 c_0 q^4}{\rho}\left(\frac{\lm^2 p_0^2}{\Gamma} + \mu\right) + \frac{q^6}{\rho}\left(\lm^2 p_0^2 K D + \Gamma K D \mu\right)=0.
    \end{split}
\end{equation}
The roots of the above equation are
\begin{align}
 \begin{split}
  \omega_1 =& \frac{ -iq^2 (\mu+\rho D+ \Gamma K \rho)}{3 \rho}+ \frac{i}{3\times  2^{1/3}\rho}(A_- -A_+) \\
  \omega_{2,3} =&\frac{-iq^2 (\mu+\rho D+ \Gamma K \rho)}{3 \rho}-\frac{i}{6\times 2^{1/3}\rho}(A_- -A_+) \pm \frac{\sqrt{3}}{6\times 2^{1/3}\rho}(A_- + A_+)
 \end{split}
 \label{eq:dispersionrelation}
\end{align}
where we have defined
\begin{align*}
 A_\pm &=\left(\sqrt{4 B^3 + C^2} \pm C \right)^{1/3}, \mathrm{~with~}\\ 
 B &=  3 \Gamma E v_1 c_0 \rho^2 q^2
 +q^4 \left(-(\mu+\rho D+ \Gamma K \rho)^2 + 3\rho K\lm^2 p_0^2 + 3\rho (\mu D + \Gamma K \mu + \Gamma K \rho D)\right), ~\mathrm{and}~ \\
 C  &=9 q^4 \rho^2  E v_1 c_0(3\lm^2 p_0^2 + \Gamma (2\mu-\rho D - \Gamma K \rho))\\
 &~~+ q^6 \left(2\mu^3 -3(D+\Gamma K)\mu^2 \rho -3 (D-\Gamma K)^2 \mu \rho^2 + 6D\Gamma K \mu \rho^2 + 3(D^3 + \Gamma^3 K^3)\rho^3 -(D+\Gamma K)^3 \rho^3 \right).
\end{align*}
In the hydrodynamic limit $(q \to 0)$, expanding $A_\pm$ gives,
\begin{equation}
 \label{eq:seriesexpand}
A_\pm = 2^{1/3} \rho \left(q \sqrt{3 \Gamma E v_1 c_0} \pm \frac{(3\lm^2 p_0^2 + \Gamma (2\mu -\rho D -\Gamma K \rho))}{2 \Gamma  } q^2 \right) + \mathcal{O}(q^3).
\end{equation}
The above expansion of $A_\pm$ is valid for 
$$q \ll \frac{2\sqrt{\Gamma^3 \rho^2 E v_1 c_0}}{\sqrt{3}\left[ 3 \lambda_-^2 p_0^2 + \Gamma 
(2\mu-\rho D -\Gamma K \rho )\right]}.$$
Substituting \eqref{eq:seriesexpand} in \eqref{eq:dispersionrelation} gives one stable mode,
\begin{equation}
\omega_1(q) = - i q^2\left(\frac{\lm^2 p_0^2}{\Gamma \rho} + \frac{\mu}{\rho }  \right)+ \mathcal{O}(q^3),
\end{equation}
and two unstable modes with equal growth rates,
\begin{align}
 \begin{split}
\omega_{2,3} =& \pm q \sqrt{\Gamma E v_1 c_0}- \frac{iq^2 (\mu+\rho D+ \Gamma K \rho)}{3 \rho}+\frac{i q^2}{6\Gamma \rho}\left(3 \lm^2 p_0^2 + \Gamma(2 \mu - \rho D -\Gamma K \rho)  \right) + \mathcal{O}(q^3)	 \\		
 =&\pm q \sqrt{\Gamma E v_1 c_0} + \frac{i q^2}{2}\left [\frac{ \lm^2 p_0^2}{\Gamma \rho} - (D+\Gamma K) \right] + \mathcal{O}(q^3).
\end{split}
\end{align}
$\omega_{2,3}(q)$ are unstable provided $\Gamma \rho (D + \Gamma K)< \lm^2 p_0^2$. It is important to note that $E$, $v_1$ do not appear in $\mathcal{O}(q^2)$ term because of cancellations, but they are crucial for the instability to survive. Further, viscosity $\mu$ also cancels out in $\mathcal{O}(q^2)$ term, thus the growth rate is independent of $\mu$ in $q \to 0$ limit. On rescaling the length by $\mathcal{L} = \mu/(\rho \sqrt{\Gamma E v_1 c_0})$ and time by $\mathcal{T} = \mu/(\rho(\Gamma E v_1 c_0))$, we obtain the following dimensionless form of the dispersion relation which does not have explicit $E$ dependence (see \cref{fig:dispersion}(b)). 

\begin{equation}\label{eq:dispersion_ND}
W_{2,3} = \pm Q  + \frac{i Q^2}{2}\left(\frac{ \lm^2 p_0^2}{\Gamma \mu} - \frac{\rho(D+\Gamma K)}{\mu}\right) + \mathcal{O}(Q^3),
\end{equation}
where $W_{2,3} = \mathcal{T}\omega_{2,3}$ and $Q = \mathcal{L} q$. In \cref{fig:dispersion}, we plot the growth rate for different $E$. 

\section{Numerical integration of the 1D equation and Pearson's correlation}
We numerically integrate the following 1D equations (also given in the main text), on a domain of length $L=10\pi$ and discretize it using $N=1024$ equispaced collocation points.
\begin{align} \label{eq:1D}
    \begin{split}
        \rho \partial_t v &= \mu \partial^2_y v + \lm p_0 E \partial^2_yc, \\
        \partial_t p &= \lm p_0 \partial_y v -\Gamma b p^3 + \Gamma K \partial^2_y p - \Gamma E \partial_y c,~\mathrm{and} \\
        \partial_t c &= -v_1 \partial_y(pc) + D \partial^2_y c.
    \end{split}
\end{align}
The time marching is performed using a second-order Adams-Bashforth scheme, and we employ a fourth-order centered finite-difference scheme for spatial derivatives. The other parameters of our simulation are $E=0.2$, $\rho=1$, $\mu=0.1$, $\Gamma = 1$, $b=1$, $D=10^{-4}$, $K=10^{-3}$, $\lambda=0.1$, $v_1=0.1$, and $c_0=1$.\\

We calculate Pearson's correlation for the velocity $v(y,t)$ and concentration field $c(y,t)$ defined as \cite{press2007}
\begin{equation}
    r(t) = \frac{\sum\limits_{i=1}^{N} \left[v(y_i,t) - \overline{v(y_i,t)}\right]\left[c(y_i,t) - \overline{c(y_i,t)}\right]}{\sqrt{\sum\limits_{i=1}^{N} \left[v(y_i,t) - \overline{v(y_i,t)}\right]^2 \sum\limits_{i=1}^{N}\left[c(y_i,t) - \overline{c(y_i,t)}\right]^2}},
\end{equation}
where the subscript $i$ denotes the value of the field at a collocation point in the simulation domain, and $\overline{(\cdot)}$ denotes spatial averaging.


\section{Dispersion for the traveling waves}
\begin{figure}[!h]
    \includegraphics[width=0.6\linewidth]{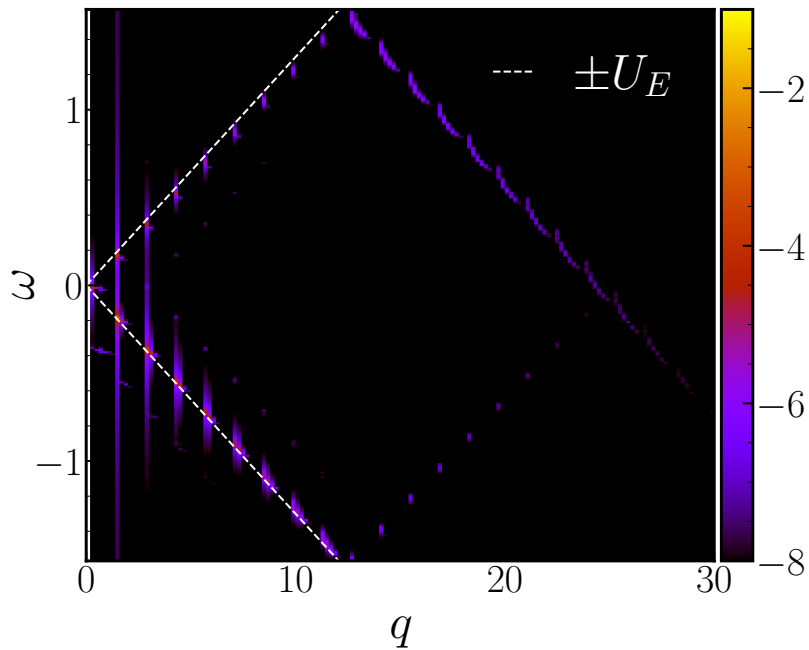}
    \caption{\label{fig:regime1} Dispersion for traveling concentration waves for $E=0.15$ in the steady state. The slopes (dashed white lines) indicates the speed of traveling waves $=\pm U_E$ where $U_E \sim \sqrt{\Gamma E v_1 c_0}$.}
\end{figure}

In \cref{fig:regime1}, we plot the dispersion $\omega(q)$ for concentration waves in the steady state for $E=0.15$. In this regime, we observe soliton-like waves passing through each other without distorting their profiles in the direction $-\bdel c/|\bdel c|$ (see Fig.~5 in the main text). The dispersion correctly captures the speed of the waves $U_E \sim  \sqrt{\Gamma E v_1 c_0}$.

\newpage
\section{Crossover regime}
\begin{figure}[!h]
    \includegraphics[width=\linewidth]{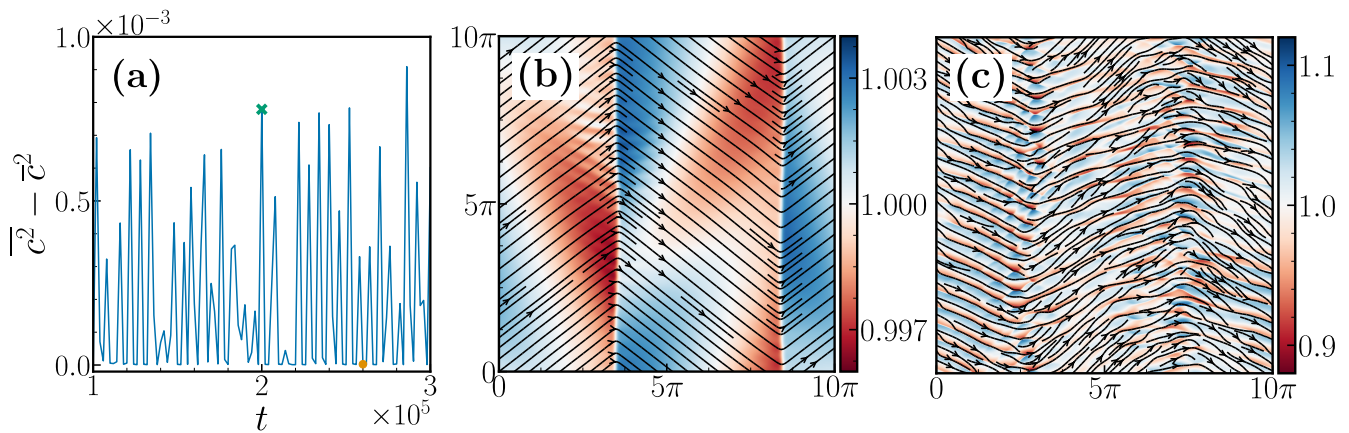}
    \caption{\label{fig:regime2} Regime II: $E=1$ (a) Time evolution of $\overline{c^2}-\overline{c}^2$ in the steady state. Pseudocolor plot of concentration overlayed with ${\bm p}$ streamlines for $\overline{c^2}=\overline{c}^2=2.3 \times 10^{-6}$ [(b), orange dot in (a)], and $\overline{c^2}-\overline{c}^2=7.8\times 10^{-4}$ [(c), green cross in (a)].}
\end{figure}

In the crossover range $0.8<E<4$ there are large fluctuations in the concentration field. In \cref{fig:regime2}, we plot the time evolution of spatial variance of the concentration field as defined in the main text and typical realizations of the order parameter and the concentration fields at a given instance in time.

\section{Correlation dimension}
\begin{figure}[h!]
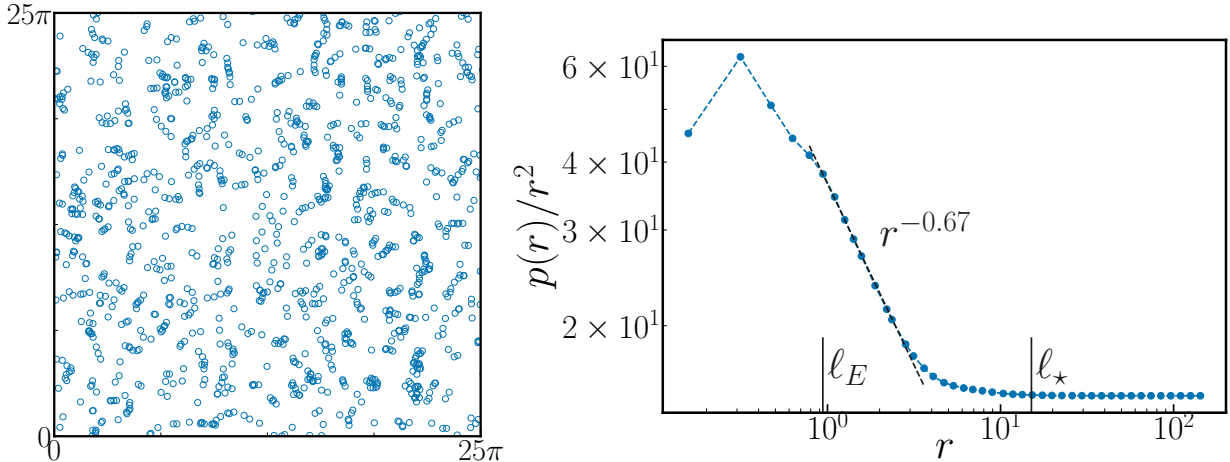

  \includegraphics[width=0.37\linewidth]{scatter_defects_E-4.pdf}
 \includegraphics[width=0.53\linewidth]{pr100pi_E-4_total.pdf}
 \caption{\label{fig:clustering}
 (Left) Scatter plot of the cores of topological defects with charge +1 showing clustering at small length scales (subdomain of $25\pi$ is shown for clarity).
 (Right) Compensated plot of the cumulative radial distribution function $p(r)$ (scaled by $r^2$) shows a scaling $\sim r^{-0.67}$ in the intermediate regime $\ell_E < r < \ell_\star$, which implies $d_2 = 1.33$.
  }
\end{figure}

Following \citet{rana2024}, we locate the cores of topological defects with charge $+1$ and analyze their spatial distribution. A uniform distribution of defects in two dimensions leads to the Porod's scaling $\Ep \sim q^{-3}$ \cite{bray2002,rana2024}. However, we find that the topological defects show signs of clustering (see \cref{fig:clustering}), which modifies the Porod's law \cite{rana2024}. To this extent, we calculate the correlation dimension $d_2$ from the spatial distribution of the defects. To find $d_2$, we compute the probability of finding two defects separated by distance $r$, $p(r) = \int_{0}^{r} s g(s) ds$, where $g(r)$ is the radial distribution function. The scaling exponent of $p(r)$ with $r$ determines $d_2$, i.e. $p(r) \sim r^{d_2}$ \cite{grassberger1983,mitra2018}. We find that $d_2 = 1.33$ for $\lE < r < \ell_\star$, which yields the modified Porod's scaling $\mathcal{E}_{\bm{p}} (q)\sim q^{-(d_2 + 1)} = q^{-2.33}$. For $r \gg \ell_\star$, $d_2 \simeq 2$.

\section{Numerical simulations with active stress}
\begin{figure}[!h]
\centering
\includegraphics[width=0.95\linewidth]{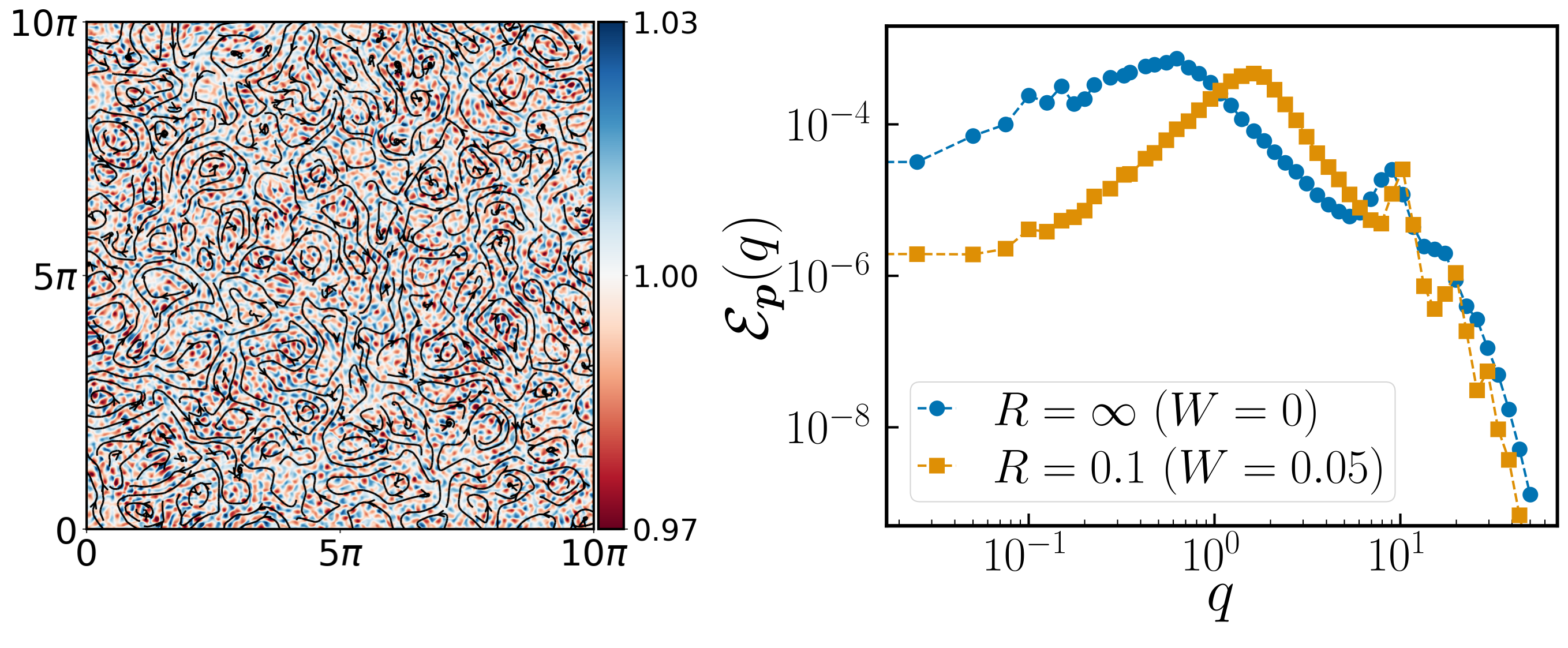}
\caption{\label{fig:R_0.1} (Left) Snapshot of concentration field overlayed with streamlines of $\pol$ shows concentration-wave turbulence for $R=0.1~(W=0.05)$ and $\Gamma E c_0/v_1=40$. Subdomain of $L=10\pi$ is shown for clarity. (Right) The order parameter spectrum for the two cases: $W=0.05$ and $W=0$ show peaks at large $q$ capturing the dominant wavenumber of concentration waves due to the novel concentration-wave instability. The traveling concentration waves for $R=0.1$ can be seen in the movie ``06\_With\_active\_stress.mp4" as described in section VIII.}
\end{figure}

We perform simulations for $R=0.1~(W=0.05)$, $\Gamma E c_0/v_1=40$, $L=80\pi$, and $N=8192$ keeping other parameters fixed to the values mentioned in the paper. We observe concentration-wave turbulence similar to $W=0$. As shown in \cref{fig:R_0.1}, the order parameter spectrum for $W=0.05$ shows a peak at large $q$ similar to $W=0$, capturing the dominant wavenumber of concentration waves due to the novel instability.

\newpage

\section{\label{sec:movies} Description of the movies}
In movies (2-6) the time  is scaled as $(t-t_0)/\mathcal{T}$, where $\mathcal{T} \equiv \mu/(\rho \Gamma E v_1 c_0)$ and the offset  $t_0$ (in the steady state) marks the beginning of the movie.  
\begin{itemize}
    \item \texttt{\textbf{01\_Instability\_mechanism\_1D.mp4}} -- Instability Mechanism\\
    The movie shows the initial time evolution of concentration and velocity fields for the 1D minimal model. It emphasizes the inertia-induced temporal lag, which is crucial for the instability. The two waves go out of phase occasionally which leads to the growth of the perturbations making the system unstable. The parameters are $L=10\pi,~N=1024$, and $E=0.2$.
    \item \texttt{\textbf{02\_Traveling\_waves\_evolution.mp4} and \textbf{03\_Traveling\_waves\_gradient\_c.mp4}} -- Traveling waves regime ($E=0.15$)\\ \texttt{\textbf{02\_Traveling\_waves\_evolution.mp4}} shows time evolution of the concentration field superimposed with streamlines of $\pol$. It shows two waves passing through each other without distorting their profiles. One of the waves is identified by $\bdel c \cdot \hat{\bm y} >0$. It moves in the direction $-\bdel c/|\bdel c|$ as shown in the movie \texttt{\textbf{03\_Traveling\_waves\_gradient\_c.mp4}}. Here, $t_0=3\times 10^5$.
    \item \texttt{\textbf{04\_Crossover\_phase.mp4}} -- Crossover phase regime ($E=1$).\\
    In this regime, the streamlines of $\pol$ are mostly ordered with spontaneous appearance and disappearance of vortical structures with a background of concentration waves. Here, $t_0=290430$.
    \item \texttt{\textbf{05\_Concentration\_wave\_turbulence.mp4}} -- Concentration-wave turbulence ($E=4$)\\
    The movie shows the evolution of the concentration field. We only plot a square subdomain of length $4 \pi$ for clarity. Traveling concentration waves similar to regime I can be seen criss-crossing. On larger scales, the streamlines of $\pol$ show complex spatiotemporal structures as seen in the $E=4$ snapshot in the main text. Here, $t_0=2 \times 10^4$.

    \item \texttt{\textbf{06\_With\_active\_stress.mp4}} -- Concentration-wave turbulence ($E=4$) in the presence of active stress, $R=0.1~(W=0.05)$. \\
   The movie shows time evolution of the concentration field. The simulations are performed on $L=80\pi$ with $N^2=8192^2$, but we only plot a square subdomain of length $4 \pi$ for clarity. Traveling concentration waves are observed similar to the $W=0$ case. Here, $t_0=12000$.
\end{itemize}

\bibliographystyle{apsrev4-2}
\bibliography{Bibliography.bib}